\documentclass[11pt]{article}
\usepackage{DGfest,epsfig}
\usepackage{times}

\def\beq{\begin{equation}}
\def\eeq{\end{equation}}
\def\bea{\begin{eqnarray}}
\def\eea{\end{eqnarray}}
\def\bq{\begin{quote}}
\def\eq{\end{quote}}

\def\bear{\begin{array}}
\def\eear{\end{array}}
\def\nnb{\nonumber}
\def\ga{\left(}
\def\dr{\right)}
\def\aga{\left\{}
\def\adr{\right\}}

\def\Lrar{\Longrightarrow}

\def\nnb{\nonumber}
\def\la{\langle}
\def\ra{\rangle}
\def\nin{\noindent}
\def\ba{\begin{array}}
\def\ea{\end{array}}

\def\mb{\overline{m}}

\def\gam5{\gamma_5}

\begin{document}
\baselineskip 11.5pt
\title{\boldmath 
%$U(1)_A $ ANOMALY,  
$U(1)_A $ TOPOLOGICAL SUSCEPTIBILITY AND ITS SLOPE, \\
PSEUDOSCALAR GLUONIUM AND THE SPIN OF THE PROTON 
% FROM QSSR
%:  \\ QCD SPECTRAL SUM RULES VERSUS LATTICE CALCULATIONS
\footnote{This review is written in honour of Professsor Adriano Di Giacomo for his 70th birthday \\ (Adrianofest 26-27th January 2006,
Pisa, Italy)}.}

\author{ Stephan NARISON}

\address{Laboratoire de Physique Th\'eorique et Astrophysiques,\\
UM2,
Place Eug\`ene Bataillon,
34095 - Montpellier Cedex 05-FR.
\\ E-mail:
snarison@yahoo.fr}

%\author{ C.D. AUTHOR2, E.F. AUTHOR3 }
%
%\address{Dipartimento di Fisica, Universit\`a di Pisa, Largo Pontecorvo 3,\\
%        I--56127 Pisa, Italy}

\maketitle\abstracts{
We review the determinations of the pseudoscalar
glueball and eventual radial excitation of the $\etaÕ$ masses and decay constants from QCD spectral sum rules 
(QSSR). The glueball mass is $(2.05\pm 0.19)$ which  one can compare with the eventual experimental candidate $X(1835)$,
while the $\eta(1400)$  is likely a radial excitation of the $\eta'$-meson.  Their effects on the estimates of $U(1)_A$ topological susceptibility and its slope as well as the impact of the latter  in the estimate of the spin of the proton is discussed. We predict the singlet polarized parton distributions to be $a^0(Q^2=4$ GeV$^2$)=$0.32\pm 0.02$, which is about a factor two smaller than the OZI value, but comparable with the COMPASS measurement of $0.24\pm 0.02$.}
% but comparable with the COMPASS measurement of $0.24\pm 0.02$.}
%%%%%%%%%%%%%%%%%%
\section*{Prologue}
%%%%%%%%%%%%%%%%%% 
\nin
It is a great honour and pleasure for me to write this review for Professor Adriano Di Giacomo for celebrating his 70th birthday.
My scientific connection with Adriano started when I studied at CERN in 82, the SVZ-expansion of QSSR \cite{SVZ,SNB}, 
and where
the Pisa group \cite{GIACO1} has presented the first quenched lattice results values of the gluon condensates
$\la\alpha_s G^2\ra$ and
$g^3f_{abc}\la G^aG^bG^c\ra$
 at the 1st Montpellier Conference on Non-pertubative methods in 85. About the same time, my previous interest 
\cite{SNG1} on the topological susceptibility
$\chi(0)$ of the $U(1)_A$ anomaly \cite{WITTEN} and its slope $\chi'(0)$ continued after several
discussions with Gabriele Veneziano during my stay at CERN . A program on the estimate of these
quantities using QSSR has started \cite{SNG2}.
In parallel, the Pisa group has runned their lattice simulations, and, for his 2nd participation at the Montpellier conference
(QCD 90), Adriano has presented the lattice Pisa group \cite{GIACO2} results for $\chi(0)$ and $\chi'(0)$  which confirmed our
previous results in \cite{SNG2}. Since then, there was a common interest between the Pisa group and Montpellier on these
non-perturbative aspects of QCD. More recently, these common interests concern the proton spin problem \cite{SHORE,SHORE2,GIACOMO3}, the estimate of the gluon condensate using QSSR \cite{SNG} and lattice \cite{GIACO4}, and the estimate of the quark-gluon mixed condensates
using QSSR\cite{SNMIX} and field correlators~\cite{MIX2}.
\\ In
94, Adriano joined the international organizing committee of the QCD-Montpellier Series of Conferences and, in 2001, the one
of the Madagascar High-Energy Physics (HEP-MAD) Series of Conferences. He is among the few committee members who send regularly
speakers and participate continuously to these conferences. The organization team of these meetings has always appreciated his
human kindness and friendship. 
%%%%%%%%%%%%
\section{Introduction}
%%%%%%%%%%%%
\nin
The $U(1)_A$ anomaly is one the most fundamental and fascinating problem of QCD, which has been solved 
in the large $N_c$ limit by 't Hooft-Witten-Veneziano \cite{WITTEN,VENEZIANO}, where Veneziano found a solution
without the uses of instantons. The $U(1)_A$ topological
susceptibility is defined as:
\beq
\chi(0)\equiv\psi_{gg}(0)\equiv i\int d^4x \la {\cal T} Q(x) Q^\dagger (0)\ra~,
\eeq
where:
\bea
Q(x)= {1\over 8\pi}\alpha_s G^{\mu\nu}_a  \tilde G_{\mu\nu}^a~, 
\eea
is the $U(1)_A$ anomaly of the singlet axial-current:
\beq
\partial ^ \mu A_\mu(x) = \sum_{q=u,d,s} \Big{[}J_q\equiv 2m_q\bar\psi_q(i\gamma_5) \psi_q\Big{]}+2n_fQ(x)~.
\eeq
where $n_f$ is the number of light quark flavours, $\psi_q$ the quark fields, $m_q$ the current quark mass and:
$
\tilde G_{\mu\nu}^a\equiv ({1/ 2})\epsilon_{\mu\nu\rho\sigma}
G^{\rho\sigma}_a~.
$
The topological charge is defined as:
\beq
Q=\int d^4x~Q(x)~,
\eeq
which is an integer for classical field configurations (2nd Chern number).
Arguments based on large $N_c$ for $SU(N_c)\times SU(n_f)$  lead to \cite{VENEZIANO} :
\beq
\chi(0)={f_\pi^2\over 2n_f}\ga M^2_{\eta'}+M^2_\eta-2M^2_K\dr .
\eeq
where, $f_\pi=(92.42\pm 0.26)$ MeV \footnote{A generalization of the result including $SU(3)$ breakings for the decay constants can be found in \cite{SHOREA}.} . For $n_f=3$:
\beq
\chi(0)\simeq \ga 180~{\rm MeV}\dr^4~.
\eeq
Gabriele Veneziano and, later on, Giancarlo Rossi have challenged us to analyze the validity of this approximation  at
finite
$N_c=3$, by computing
$\chi(0)$ and $\chi'(0)$ using QSSR, because the previous result is obtained at
$q^2=M^2_{\eta'}$ but not at $q^2=0$, where,  they have used the expansion:
\beq\label{eq: expansion}
\chi(q^2)\simeq \chi(0)+q^2\chi'(0)+...,
\eeq
%%%%%%%%%%%%
\section{The pseudoscalar gluonium/glueball mass from QSSR}
%%%%%%%%%%%%
QSSR has been used to calculate the topological charge in pure Yang-Mills QCD. In so doing, 
one has first extracted the mass and decay constant of the pseudoscalar glueball/gluonium using either a numerical fitting
procedure \cite{SNG1,SNG2} or sum rule optimization criteria \cite{SNG3,SNG4} from the unsubtracted sum rules (USR):
\beq
{\cal L}(\tau)=\int_0^{t_c}dt e^{-t\tau}{\rm Im}~\chi(t)~~~~~~{\rm and}~~~~~~{\cal R}(\tau)=-{d\over d\tau}\log {\cal L}(\tau) ~,
\eeq 
where the QCD model coming from the discontinuity of the QCD graphs has been used  for the continuum. The analysis gives:
\beq
M_G =  (2.05\pm 0.19) ~{\rm GeV} ~ ~~~{\rm and}~ ~~~ f_{G}= (8-17)~{\rm MeV}~,
\eeq
corresponding to $t_c=(6-7)$ GeV$^2$, and where $f_G$ is normalized as:
\beq
\la 0|Q(x)|G\ra=\sqrt{2}f_GM^2_G~,
\eeq
i.e. like $f_\pi/(2n_f)$ with $f_\pi=(92.42\pm 0.26)$$(92.42\pm 0.26)$ MeV.  The positivity of the spectral function gives:
\beq
M_G \leq  (2.34\pm 0.42) ~{\rm GeV}~.
\eeq
One can compare the previous values with the quenched lattice results \cite{LATT1}:
\beq
M_G =  (2.1-2.5)         ~{\rm GeV}~. 
\eeq
Recent sum rule analysis using instanton liquid model leads to similar results: $M_G\simeq 2.2$ GeV and $f_G\simeq 17$ MeV
in our normalization \cite{FORKEL}, though the same approach leads to some inconsistencies in the scalar channel \cite{SNS}. One
should note that in the previous analysis~\cite{SNG3,SNG4} and \cite{FORKEL}, the value of the gluonium decay constant is
smaller than the corresponding value of the
$\eta'$ one which is about
24 MeV in the chiral limit \cite{SNG1,SNG2,SHORE} (see section \ref{sec: massless}). In a recent paper \cite{HE}, the mass of the eventual experimental pseudoscalar gluonium
candidate $X(1835)$ has been used as input and the Laplace sum rule ${\cal L}$ has been exploited for fixing the corresponding
decay constant. Though (a priori) self-consistent, this analysis is less constrained than the previous sum rules used  in
\cite{SNS,FORKEL}, where the two sum rules ${\cal L}$ and ${\cal R}$ have been simultaneously used for constraining the gluonium
mass and decay constant. The resulting value of the gluonium decay constant is about (8-12) MeV \cite{HE} and agrees with the previous values. On the other, a $G$-$\eta_c$ mixing due to direct instanton has been considered in
\cite{KOCHELEV} for pushing the unmixed gluonium mass of about 2.1 GeV down to 1.8 GeV with a mixing angle of about 17$^0$.This value is comparable  with the early value  of about $\theta_P=12^0$ of the meson-gluonium mixing angle using  the OPE of  the off-diagonal light quark-gluon correlator:
\beq\label{eq:off}
\psi_{gq}(q^2)\equiv {i\over 2n_f}\int d^4x e^{iqx}\la {\cal T} Q(x) J^\dagger_q(0)\ra~,
\eeq
which is proportional to $m_s$ \cite{PAK}.  Indeed, it is also plausible that the $X(1845)$ comes from a complicated  mixing of the glueball with the $\eta(1440)$ and $\eta_c$. A confirmation of the gluonium nature of the $X(1835)$ requires some more independent tests.
%%%%%%%%%%%%%%%%%%%%%%%%%%%%%%%%%%
\section{On the natures of the $\eta(1295)$ and $\eta(1400)$ from QSSR }
There are two other experimental candidates which are the $\eta(1295)$ and $\eta(1400)$ \cite{PDG}, which one can intuitively interpret as the first 
radial excitations of the $\eta(547)$ and $\eta'(958)$. Here, we shall test if the $\eta(1400)$ satisfies this interpretation.
Using the same approach, the nature of the $\eta(1400)$ \cite{PDG}  has been 
tested by measuring its coupling to the $U(1)_A$ singlet  current $Q(x)$ in the chiral limit. In this case, we can work with the SSR:
\beq
\int_0^{t_c}{dt\over t}~e^{-t\tau}{1\over\pi}{\rm Im}~\chi(t)~,
\eeq
as one expects from large $N_c$ arguments that the 
topological charge $\chi(0)$ vanishes due to the $\theta$-independence of the QCD Lagrangian, thanks to the presence of
the singlet $\eta_1$.
Including in  the sum rule, the contributions of the $\eta'$, $\eta(1400)$, $G$ and the QCD continuum  with $t_c$ fixed previously,  one cannot find any room to put the $\eta(1400)$, i.e. $f_\eta\approx 0.$ Relaxing the constraint by replacing the QCD continuum with the $\eta(1400)$, one can deduce:
\beq
f_{\eta(1400)}\leq 16 ~{\rm MeV} \ll f_{\eta'}\simeq  (24.1\pm 3.5)~{\rm MeV}~,
 \eeq
which should be a weak upper bound within this assumption. This feature indicates that the $\eta(1400)$ is likely the first radial excitation of the $\eta'$, as intuitively expected, while the glueball has a higher mass.  
%%%%%%%%%%%%%%%%%%%%%%%%%%%%%%%%%%%%%%%%%%%%%%%%%%%%%%%%%%%%
\section{The \boldmath $U(1)_A$ topological susceptibility from QSSR}
%%%%%%%%%%%%%%%%%%%%%%%%%%%%%%%%%%%%%%%%%%%%%%%%%%%%%%%%%%%%
\subsection{The pure Yang-Mills result}
Applying the Borel/Laplace  operator to the subtracted quantity:
\beq
{\chi(q^2)-\chi(0)\vert_{\rm no~quarks}\over q^2}~,
\eeq
one can derive a combination of the unsubtracted sum rule (USR) and subtracted sum rule (SSR) for the topological
suceptibility in  pure Yang-Mills without quarks \cite{SNG1,SNG2}:
\bea\label{eq: chi}
\chi(0)\vert_{\rm no~quarks}&=&\int_0^{t_c}{dt \over  t}e^{-t\tau}\ga 1
-{t\tau\over 2}\dr{1\over\pi}{\rm Im}~\chi(t)-\nnb\\
&&\ga {\alpha_s\over 8\pi}\dr^2{2\over \pi^2}\tau^{-2}\aga {1\over\log{\tau\Lambda^2}}+2\pi^2\la G^2\ra+6\pi^2g\la
G^3\ra\tau^3\adr~.
\eea
The value of the gluon $\la \alpha_s G^2\ra=(0.07\pm 0.01)$ GeV$^2$ from $e^+e^-$ data and heavy-quark mass-splitting \cite{SNG,SNB}
has been confirmed by lattice calculations with dynamical fermions \cite{GIACO4}, while the value of the triple gluon condensate
$g^3f_{abc}\la G^aG^bG^c\ra$ is about 1.5 GeV$^2\la \alpha_s G^2\ra$ from instanton model \cite{NSVZ} and lattice calculations \cite{GIACO1}. 
The previous combination of sum rules is more interesting than the alone SSR, as the effect of the QCD continuum is minimized here.
At the sum rule optimization scale of $\tau$ about $0.5$ GeV$^{-2}$, it leads to the value \cite{SNG1,SNG2}:
\beq
\chi(0)\vert_{\rm no~quarks}\simeq -[(106-122)~{\rm MeV}]^4\approx -4\ga{\alpha_s\over 8\pi}\dr^2 \la G^2\ra~,
\eeq
indicating the role of the gluon condensate in the determination of $\chi(0)$. The sign and the size of this result, though
inaccurate are in agreement with the large $N_c$ results obtained previously. 
In pure gauge,  lattice gives the value \cite{GIACO2,GIACO5,GIUSTI}:
\beq
\chi(0)\vert_{SU(2)}\simeq -[(167\pm 25)~{\rm MeV}]^4~,~~~~~~~~~~\chi(0)\vert_{SU(3)}\simeq -[(191\pm 5)~{\rm MeV}]^4~,
\eeq
which one can compare  with the two former ones from large $N_c$ and from the sum rules.
%%%%%%%%%%%%%%%%%%%%%%%%%%%%%%%%%%%%%%%%%%%%%%%%%%%%%%%%%%%%%%%%%%%%%%%%%%%%%%%%%%%%
\subsection{Result in the presence of quarks}
%%%%%%%%%%%%%%%%%%%%%%%%%%%%%%%%%%%%%%%%%%%%%%%%%%%%%%%%%%%%%%%%%%%%%%%%%%%%%%%%%%%%
\nin
In this case the analysis is more involved as one also has to consider the diagonal quark-quark correlator \footnote{These correlators
include in their definitions the quark mass through $J_q$, while in \cite{SHORE2}, this quark mass is factorized out.}:
\beq\label{eq: qq}
\psi_{qq}(q^2)\equiv {i \over( 2n_f)^2}\int d^4x~ e^{iqx}\la {\cal T} J_q(x) J^\dagger_q(0)\ra~,
\eeq
 and off-diagonal quark-gluon correlator in Eq. (\ref{eq:off}). Then, the full correlator reads:
\beq\label{eq: full}
\psi_5(q^2)=\psi_{gg}(q^2)+2\psi_{gq}(q^2)+\psi_{qq}(q^2)~.
\eeq
In this case, $\psi_5(0)$ is not vanishing and can be deduced from chiral Ward identities to be~\cite{VENEZIANO,SHORE2}:
\beq
 \psi_5(0)=-{4\over (2n_f)^2}\sum_{q=u,d,s}m_q\la\bar qq\ra~,
\eeq
Therefore,  the 
topological charge $\chi(0)$ vanishes in the chiral limit due to the $\theta$ independence of the QCD Lagrangian. 
Lattice calculations in full QCD for two degenerate dynamical fermions find \cite{GIACO6}:
\beq
\psi_5(0)=[(163\pm 6)~{\rm MeV}]^4~,
\eeq
in agreement with the large $N_c$ result \cite{WITTEN,VENEZIANO} but does not have enough accuracy for checking the
linear $m_q$-dependence expected from current algebra.\\
In this case of massive quarks, 
we include $SU(3)$ breakings to the sum rule in order to see the effect of the physical $\eta'$. 
Including both the gluonium and QCD continuum contribution to the appropriate sum rule, one can see
from \cite{SNG1,SNG2} that their contributions tend to compensate, and then  give the approximate numerical formula
to leading order:
\beq
2M^2_{\eta'}f^2_{\eta'}e^{-M^2_{\eta'}\tau}\ga 1-{M^2_{\eta'}\tau\over 2}\dr\simeq
\psi_5(0)+{3\over 4\pi^2}\ga{\mb_s\over 2n_f}\dr^2\tau^{-1}~.
\eeq
Using the quark model prediction:
\beq\label{eq: fpi}
f_{\eta'}\simeq {1\over 2n_f}\sqrt{3}f_\pi\simeq 27~{\rm MeV}~,
\eeq
 the physical $\eta'$ mass and the value $\mb_s(\tau)\simeq 100$ MeV \cite{SNMS}, we can deduce at the sum rule optimization scale $\tau\simeq 0.5~{\rm GeV}^{-2}$:
\beq
\psi_5(0)\simeq [157~{\rm MeV}]^4~,
\eeq
in good agreement with the large $N_c$ and lattice results, then showing the consistency of the sum rule approach.
Armed with these consistency checks, we shall now evaluate with confidence the slope of the topological susceptibility
$\chi'(0)$.
%%%%%%%%%%%%%%%%%%%%%%%%%%%%%%%%%%%%%%%%%%%%%%%%%%%%%%%%%%%%%%%%%%%%%%%%
\section{The slope \boldmath $\chi'(0)$ of the topological susceptibility from QSSR} 
Using the $q^2$ expansion in Eq. (\ref{eq: expansion}),
one can derive a sum rule for $\chi'(0)$ by applying, the Laplace sum rule operator, to the twice subtracted quantity:
\beq
{\chi(q^2)-\chi(0)-q^2\chi'(0)\over (q^2)^2}=\int{dt\over t}{1\over t-q^2-i\epsilon}{1\over\pi}{\rm Im}~\chi(t)~,
\eeq

%%%%%%%%%%%%%%%%%%%%%%%%%%%%%%%%%%%%%
\subsection{The pure Yang-Mills case}
One obtains in this case:
\beq
\int_0^\infty{dt\over t^2}e^{-t\tau}{1\over\pi}{\rm Im}~\chi(t)={\cal F}_1(\tau)-\chi(0)\tau+\chi'(0)~,
\eeq
where ${\cal F}_1(\tau)$ comes from the OPE expression of the correlator $\chi(q^2)$ \cite{SNG1}. Eliminating $\chi(0)$
by using Eq.~(\ref{eq: chi}), one can deduce the sum rule \cite{SNG2}:
\bea
&&\int_0^\infty{dt\over t^2}e^{-t\tau}\Big{[} 1-t\tau\ga 1-{t\tau\over 2}\dr{1\over\pi}{\rm Im}~\chi(t)\Big{]}
\simeq \chi'(0)+\ga {\alpha_s\over 8\pi}\dr^2{2\over \pi^2}\tau^{-1}\times\nnb\\
&&\aga {1+
\ga{\alpha_s\over\pi}\dr\Big{[} 10-2\gamma_E\beta_1-2{\beta_2\over\beta_1}\log\ga -\log{\tau\Lambda^2}\dr\Big{]}+4\pi^2g\la
G^3\ra\tau^3}\adr~,
\eea
where $\beta_1=-11/2$ and $\beta_2=-51/4$ for pure gauge $SU(3)_c$. Using $\tau$- and $t_c$-stability criteria, and
saturating the spectral function by the pseudoscalar glueball and QCD continuum, one can deduce at $\tau\simeq 0.5$
GeV$^{-2}$ \cite{SNG2}:
\beq
\chi'(0)\vert_{\rm no~quarks}\simeq -[7\pm 3)~{\rm MeV}]^2~,
\eeq
indicating that:
$
\chi'(0)M_{\eta'}^2\ll \chi(0)~,
$
which justifies the accuracy of the large $N_c$ result. 
This value and the sign has been confirmed by lattice calculations~\cite{GIACO2}:
\beq
\chi'(0)\vert_{\rm quenched}\simeq -\big{[}(9.8\pm 0.9)~{\rm MeV}\big{]}^2~.
\eeq
%%%%%%%%%%%%%%%%%%%%%%%%%%%%%%%%%%%%%%%%%%%%%%
\subsection{Result in the presence of massless quarks}\label{sec: massless}
In this case, quark loops enter into the value of the $\beta$ function and of the QCD scale $\Lambda$. The $\eta'$ enters now into
the spectral function. Its coupling to the gluonic current can be estimated from the SSR:
\beq
\int_0^{t_c}{dt \over  t}e^{-t\tau}\ga 1
-{t\tau\over 2}\dr{1\over\pi}{\rm Im}~\chi(t)~,
\eeq
taking into account the fact that in the chiral limit $\chi(0)=0$. Therefore, one obtains \cite{SHORE}:
\beq\label{eq: feta}
f_{\eta'}\ga\tau\simeq 0.5 ~{\rm GeV^{-2}}\dr\simeq (24.1\pm 3.5)~{\rm MeV}~,
\eeq
where the decay constant has a weak $\tau$-dependence because of renormalization:
\beq
f_{\eta'}(\tau)\simeq \hat f_{\eta'} {\rm exp}\ga {8\over -\beta_1^2\log {\tau \Lambda^2}}\dr~,
\eeq
where $\hat f_{\eta'}$ is RG invariant. This result can be compared with the quark model prediction in Eq. (\ref{eq: fpi}).
Using the Laplace and FESR versions of the twice subtracted sum rule, one finds~\cite{SHORE}:
\beq\label{eq: psiprim1}
\sqrt{\chi'(0)}\vert_{\rm massless}(\tau)\simeq \big{[}(26.5\pm 3.1)~{\rm MeV}\big{]}^2~.
\eeq
The result has changed sign compared to the one from pure Yang-Mills, while its absolute value is about 12 times higher. The main effect is due to the $\eta'$ which gives the dominant contribution to the spectral function 
compared to the
gluonium and QCD continuum. That can be understood because of its low mass and of the fact that its decay
constant $f_{\eta'}$ is larger than the gluonium decay constant $f_G$. \\
Similar estimate of 
$\psi_5(0)$ in the chiral limit for the singlet channel has been done in \cite{INDIAN}, where the value of $\chi(0)'$ is significantly larger by
about a factor 2.5 than ours. However, we found some inconsistencies in the two sides of the SR: in the experimental side the
contributions of the singlet and octet mesons have been considered and the phenomenological value of the decays constants, masses and mixing angles 
have been used; while in the QCD side, only the correlator associated to the singlet current $Q(x)$, or to massless quark has been accounted for. \\
On the other, some criticisms raised in
a series of paper~\cite{IOFFE}, due to direct instanton breaking of the OPE, do not apply in our case: our different sum rules optimize at a large scale $\tau\leq
0.5$ GeV$^{-2}$, where this contribution become irrelevant like other high-dimension condensates. In \cite{FORKEL}, it has been argued that
screening corrections cancel the direct instanton contributions, indicating that these effects are not yet well understood. However, in your approach, the introduction of the new $1/q^2$-term in the OPE, which is also negligible, is an alternative to the direct instanton contributions
\cite{ZAK,CNZ}. Some answers to these criticisms have been already explicitly given in \cite{SHORE2}.
%%%%%%%%%%%%%%%%%%%%%%%%%%%%%%%%%%%%%%%%%%%%%%
\subsection{Result in the presence of massive quarks}
As can be seen from the expression of the two-point correlator $\psi_5(q^2)$ in Eq. (\ref{eq: full}), the analysis is more involved.
In this case, its slope has been estimated from the substracted Laplace sum rules. At the stability point $\tau\simeq (0.2-0.4)$
GeV$^{-2}$ , one finds \cite{SHORE2}:
\beq\label{eq: psiprim2}
\sqrt{\psi'_5(0)}= (33.5\pm 3.9)~{\rm MeV}~,
\eeq
while the $\eta'$-decay constant becomes:
\beq
f_{\eta'}\ga\tau\dr\simeq (27.4\pm 3.7)~{\rm MeV}~,
\eeq
compared with the results for massless quarks in Eqs. (\ref{eq: psiprim1}) and (\ref{eq: feta}) and agrees quite well with the quark model prediction in Eq. (\ref{eq: fpi}). One can notice that the $SU(3)$ breaking  
effect is about 10\% and 20\% respectively showing a smooth dependence on the strange quark mass as expected.
Similar analysis can be done in the flavour non-singlet case by working with the correlator associated to the $\eta$-meson current:
\beq
\la 0|\partial^\mu J^8_{\mu 5}|\eta\ra=f_\eta M^2_\eta~,
\eeq
where in the $SU(3)$ limit and in this normalization without $1/2n_f$, $f_\eta=f_\pi=(92.42\pm 0.26)$ MeV. Therefore, one obtains \cite{SHORE2}:
\beq\label{eq: psiprim3}
\sqrt{\psi_5^{'88}(0)}= (43.8\pm 5.0)~{\rm MeV}~,~~~~~{\rm and}~~~~~f_\eta/f_\pi=1.37\pm 0.16~.
\eeq
Instead of the value of $f_\pi^{\rm exp}=(92.42\pm 0.26)$ MeV, we have used the sum rule prediction \cite{SHORE2}:
\beq
f_\pi=(107\pm 12)~{\rm MeV}~,
\eeq
 for a self-consistency of the whole results. The value of the $SU(3)$ breaking ratio $f_\eta/f_\pi$ is
in line with $f_K/f_\pi=1.2$, where we expect bigger effects for the $\eta$ than for the $K$. 
%%%%%%%%%%%%%%%%%%%%%%%%%%%%%%%%%%%%%%%%%
\section{Applications of the QSSR results to the proton spin}
%%%%%%%%%%%%%%%%%%%%%%%%%%%%%%%%%%%%%%%%%
The previous results have been applied to the proton spin problem, where one expects that the gluon content of the proton is due to the
$U(1)_A$ anomaly. This property can be made explicit in the approach of DIS where the matrix elements from the OPE are factorised into
composite operators and proper vertex functions \cite{SHORE3}. In the case of polarised $\mu p$ scattering, the composite operator can be
identified with the slope $\chi'(0)$ of the topological susceptibility, which is an universal quantity and then target independent, while the
corresponding proper vertex is renormalisation group invariant. The first moment of the polarised structure function reads, in terms of
the axial charges of the proton:
\beq\label{eq: g1}
\Gamma_1^p(Q^2)\equiv \int_0^1 dx~g_1^p(x,Q^2)={1\over 12}C_1^{NS}(\alpha_s(Q^2))\ga a^3+{1\over 3}a^8\dr+{1\over 9}C_1^S(\alpha_s(Q^2))a^0(Q^2)~,
\eeq
where the Wilson coefficients arise from the OPE of the two electromagnetic currents:  
\beq
C_1^{NS}=1-a_s-3.583a_s^2-20.215a_s^3~,~~~~~~~~~~~~~~~~~~C_1^{S}=1-{1\over 3}a_s-0.550a_s^2~,
\eeq
where $a_s\equiv \alpha_s/\pi$. 
The axial charge are defined from the forward matrix elements as:
\beq
\la p,s|J^3_{\mu 5}|p,s\ra={1\over 2}a^3s_\mu~,~~~ \la p,s|J^8_{\mu 5}|p,s\ra={1\over 2\sqrt 3}a^8s_\mu~,~~~ \la p,s|J^0_{\mu
5}|p,s\ra=a^0(Q^2)s_\mu~,
\eeq
where $J^a_{\mu 5}$ are the axial currents and $s_\mu$ the proton polarisation vector. 
Using QCD parton model, the axial charges read, in terms of moments of parton distributions  \cite{ALTARELLI}:
\beq
a^3=\Delta u-\Delta d~,~~~a^8=\Delta u+\Delta d- 2\Delta s~,~~~ a^0=[ \Delta \Sigma\equiv  \Delta u+\Delta d+ \Delta s]-{n_f\over 2}{a_s}\Delta g(Q^2)~.
\eeq
$a^3$ and $a^8$ are known in terms of the $F$ and $D$
coefficients from beta and hyperon decays:
\beq
a^3=F+D~~~~~~~{\rm and} ~~~~~~~a^8=3F-D~,
\eeq
where \cite{BOURQUIN}:
\beq
F+D=1.257\pm 0.008 ~~~~~~~{\rm and} ~~~~~~~F/D=0.575\pm 0.016~,
\eeq
so that an experimental determination of the first moment of $g_1^p$ in polarised deep inelastic
scattering (DIS) allows a determination of the singlet axial charge $a^0(Q^2)$. In the na\"\i ve quark or valence quark, parton model, one expects that $\Delta s=\Delta g=0$, and then $a^0=a^8$, which is the OZI prediction.
The {\it proton spin problem} is that the experimental value of
{\it $a_0(Q^2)$ is much smaller than $a^8$}, which would be its expected value if the OZI rule were exact in this channel.
The first SMC data found  \cite{SMC,SMC2} :
\beq
a^0(Q^2=10~{\rm GeV}^2)=0.19\pm 0.17~ ~~~~~~~{\rm and} ~~~~~~~ a^0(Q^2=5~{\rm GeV}^2)=0.19\pm 0.06 ~,
\eeq
which are confirmed and improved by the recent COMPASS data at $Q^2=4$ GeV$^2$ \cite{COMPASS}:
\beq\label{eq: compass}
a^0(Q^2=4~{\rm GeV}^2)=0.24\pm 0.02~.
\eeq
These results are much smaller than the OZI prediction ($\Delta g=0$):
\beq\label{eq: ozi}
\Delta\Sigma\vert_{\rm OZI}=3F-D=0.579\pm 0.021~.
\eeq
Inserted into Eq. (\ref{eq: g1}), the OZI results lead to the Ellis-Jaffe sum rule \cite{ELLIS}. It has been conjectured in
\cite{ALTARELLI} that  the suppression of $a^0$ with respect to $a^8$ is due to the gluon distribution.  However, the SMC measurement
of $\Delta \Sigma$ gives a value \cite{SMC2}:
\beq
\Delta\Sigma\vert_{\rm SMC}\simeq 0.38\pm 0.04~,
\eeq
indicating that independently of $\Delta g$, there is also a large difference between the OZI prediction and the data \footnote{In \cite{DOSCH3}, an estimate of the scalar sea quark content of the nucleon indicates that it is suppressed compared to the valence one. Similar results may apply here.}. In the following, we will not try to solve this discrepancy but will show our prediction for $a^0$, where the sum of the quark and gluon components are concerned.
%%%%%%%%%%%%%%%%%%%%%%%%%%%%%%%%%%%%%%%%%%%
\subsection{Prediction in the chiral limit}
Using the composite operator $\oplus$ proper vertex functions approach
\cite{SHORE}, one can write the decomposition, in the chiral limit:
\beq
\Gamma_{1~\rm singlet}^p={1\over 9}{1\over 2M_N}2n_fC^S_1(\alpha_s(Q^2))\sqrt{\chi'(0)}\vert_{Q^2}\hat\Gamma_{\eta^0NN}~.
\eeq
The key assumption is that the vertex is well approximated by its OZI value:
\beq
\hat\Gamma_{\eta^0NN}=\sqrt{2}\hat\Gamma_{\eta^8NN}~,
\eeq  
while all OZI violation  in $\Gamma_{1~\rm singlet}^p$ is contained inside $\sqrt{\chi'(0)}$. Comparing the result with the OZI prediction of $a^8$,
one can deduce:
\beq\label{eq: ratio1}
{a^0(Q^2)\over a^8}={\sqrt{6}\over f_\pi}\sqrt{\chi'(0)}\vert_{Q^2}= 0.60\pm 0.12~,
\eeq
which gives our original {\it proton spin sum rule} \cite{SHORE}:
\beq
a^0(Q^2=10~{\rm GeV}^2)=0.35\pm 0.05~~~\Lrar~~~ \Gamma^p_1(Q^2=10~{\rm GeV}^2)=0.143\pm 0.005~,
\eeq
where we have used:
\beq
\sqrt{\chi'(0)}\vert_{Q^2=10~{\rm GeV}^2}=(23.2\pm 2.4)~{\rm MeV}~,
\eeq
after running the result in Eq. (\ref{eq: psiprim1}) from 2 to 10 GeV$^2$. These results can be compared with the OZI value in 
Eq. (\ref{eq: ozi}) and agree with the last SMC data at $Q^2=10~{\rm GeV}^2$ \cite{SMC2}:
\beq
\Gamma^p_1(Q^2)=0.145\pm 0.008\pm 0.011~~\Lrar~~a^0(Q^2)=0.37\pm 0.07\pm 0.10~,
\eeq
%%%%%%%%%%%%%%%%%%%%%%%%%%%%%%%%%%%%%%%
\subsection{The case of massive quarks}
In this case, the relation in Eq. (\ref{eq: ratio1}) is replaced by \cite{SHORE2}:
\beq
{a^0(Q^2)\over a^8}={1\over \sqrt{2}}{\sqrt{\psi_5'(0)}\over \sqrt{\psi'^{88}_5(0)}}=0.55\pm 0.02~,
\eeq
where $\psi_5(q^2)$ has been defined in Eq. (\ref{eq: full}) and $\psi_5^{88}(q^2)$ is the two-point correlator for the octet current.
The running of the subtraction constant $\psi_5'(0)$ is very smooth from $\tau^{-1}$= 3 to 10 GeV$^2$. Using $\psi'^{88}(0)\equiv a^8=0.58$, the previous relation leads to:
\beq
a^0(Q^2=4~{\rm GeV}^2)\simeq a^0(Q^2=10~{\rm GeV}^2)\simeq 0.32\pm 0.02~,
\eeq
which is almost equal to its value in the chiral limit. This prediction is comparable with the COMPASS result 
$(0.24\pm 0.02)$ \cite{COMPASS} given in Eq. (\ref{eq: compass}) but is about a factor 2 smaller than its OZI value in Eq. (\ref{eq: ozi}).
Using this result, we predict:
\beq
\Gamma^p_1(Q^2=4~{\rm GeV}^2)=0.145\pm 0.002~,
\eeq
where we have used the value of $\alpha_s(M_\tau)\simeq 0.347\pm 0.03$ \cite{BNP,EXP,PDG}, and the previous values of F and D from
\cite{BOURQUIN}. Improvements of these results at COMPASS energy require a good control of the higher twist corrections in the PT
expressions of $\Gamma^p_1$ which are expected to be bigger here than at SMC. Using  the sum rule estimate in \cite{ROSS}   of about $-0.03$ of the coefficient of the 1st power $1/Q^2$ corrections, one  obtains  a correction of about -$(0.008\pm 0.008)$ to $\Gamma^p_1$ at 4 GeV$^2$, where the error is an educated guess taking into account larger absolute value obtained from the fit of the Bjorken sum rule in \cite{SHORE}.
A control of the corrections in the assumption of the validity of OZI for the singlet and non-singlet vertex functions is not yet testable \footnote{A QSSR evaluation of the $\eta NN$ coupling has not been conclusive \cite{PAVER2}.}. Including the power correction, we consider as a final estimate:
\beq
\Gamma^p_1(Q^2=4~{\rm GeV}^2)=0.137\pm 0.008~.
\eeq
 Several tests of the approach have been also proposed in the literature \cite{SHORE4}. Another crucial test of our result will be a lattice measurement of $\psi'_5(0)$ with
massive dynamical fermions, which we wish that the Pisa group puts in its agenda. Unfortunately, an attempt to measure $a^0(Q^2)$ on the
lattice
\cite{GIACOMO3} has not been conclusive, as it gives a value:
\beq
a^0(Q^2)=0.04\pm 0.04\pm 0.20~,
\eeq
where the last error is an educated guess of the effective error expected by the authors.
%%%%%%%%%%%%%
\section{Conclusions} 
We have used QSSR for predicting the gluonium decay constant and  mass, and for arguing  that the $\eta(1440)$ is likely the radial excitation of the $\eta'(958)$. These results have been used for predicting  the topological susceptibility and its slope, which are useful inputs in the resolution of the ``proton spin problem'' .
%%%%%%%%%%%%%%%%%%%%%%%%%%%%
\section*{Acknowledgements}
%%%%%%%%%%%%%%%%%%%%%%%%%%%%
Previous collaborations with Graham Shore and Gabriele Veneziano have lead to most parts of this review. 
Communications with Graham Shore are also acknowledged. I wish to thank the organizers of the Adrianofest for inviting me to
contribute to it. 
%%%%%%%%%%%%%%%%%%%%%%%%%%%%
\section*{References}
%%%%%%%%%%%%%%%%%%%%%%%%%%%%
%%%%%%%%%%%%%%%%%%%%%%%%%%%

\end{document}